\documentclass[12pt]{article}
\usepackage[margin=1in,letterpaper]{geometry}
\usepackage{graphicx}
\usepackage{rotating}
\graphicspath{%
    {converted_graphics/}% inserted by PCTeX
    {/}% inserted by PCTeX
}
\begin{document}

\newcommand{\comment}[1]{}

\title{Self-organization of signal transduction}
\author{Gabriele Scheler\\
Carl Correns Foundation for Mathematical Biology\\
1030 Judson Drive\\
Mountain View, Ca. 94040\\
%scheler@theoretical-biology.org
}
\date{}
\maketitle
\begin{abstract}
We propose a model of parameter learning for signal transduction, where the objective function is defined by signal transmission efficiency. We apply this to learn kinetic rates as a form of evolutionary learning, and look for parameters which satisfy the objective. This is a novel approach compared to the usual technique of adjusting parameters only on the basis of experimental data. 
The resulting model is self-organizing, i.e. perturbations in protein concentrations or changes in extracellular signaling will automatically lead to adaptation.
We systematically perturb protein concentrations and observe the response of the system. We find compensatory or co-regulation of protein expression levels.
In a novel experiment, we alter the distribution of extracellular signaling, and observe adaptation based on optimizing signal transmission. 
We also discuss the relationship between signaling with and without transients. Signaling by transients may involve maximization of signal transmission efficiency for the peak response, but a minimization in steady-state responses. With an appropriate objective function, this can also be achieved by concentration adjustment.
Self-organizing systems may be predictive of unwanted drug interference effects, since they aim to mimic complex cellular adaptation in a unified way.
\end{abstract}

\section{Introduction}
%\vspace*{-0.2cm}
Signal transduction systems are often modeled as networks of biochemical kinetic equations implemented as continuous-time dynamical models using differential equations
 \cite{Bhalla99,Hoops2006}. 
If we regard a subset of species as inputs, and make sure that the system always converges to equilibrium values by using weakly reversible 
equations \cite{Deng2011,Akle2011,vanderSchaft2011}, we may transform these models 
into a set of matrices fulfilling the role of input-output transfer functions, i.e. a mapping 
from sustained input signal levels to steady-state concentrations for all target 
species \cite{Scheler2012}. Protein signaling functions (psfs) are a systemic generalization of individual dose-response functions, which are usually described by Hill equations \cite{Barlow89}. In contrast to Hill equations, which are not available for enzymatic reactions, which only calculate relative concentrations, and which only work for one reaction in isolation, the psf system calculates enzymatic and complex formation reactions in a complex systemic environment using absolute 
concentrations \cite{Scheler2012}. In addition, the reaction times to equilibrium are calculated as delay values, and the dynamic shape ('transients') is also available for further analysis.

In this paper, we want to ask the question of optimality of signal transduction. From an evolutionary standpoint, we assume that any biological signal transduction system is constructed with optimized efficiency of signal transmission. Furthermore, we assume that cells have the ability to adapt to perturbations of protein concentrations and changes in extracellular signaling by reinstating signal transmission efficiency.  In the following, we investigate this question using a biologically realistic system - beta-adrenergic signaling in a submembrane compartment of a mouse embryonic fibroblast- for a single input scenario, focusing on a selected target species as relevant output or actuator of the system (Fig.~\ref{fig1a}~A). 

Experimental analysis of signal transduction systems has shown fold-change responses to changes in input \cite{Ferrell2009,Buijsman2012}. 
Accordingly, input-output transfer functions usually follow the shape of hyperbolic (saturating) curves, which are equivalent to sigmoids for logarithmically scaled 
input \cite{Scheler2012}. 
 In Fig.~\ref{fig1a}~B
we show the effect of a knock-out (KO) for a RGS protein % \cite{Haoetal2003} 
in an experimental assay in 
yeast \cite{Yi2003}, and compare this with the effect in the model system \cite{Blackman2011}. We see 
that the effects of the RGS KO on dose-response signaling efficiency are robust across very different cellular systems. 

Usually, when we use a computational model to investigate perturbations, we only study the effects as reflected in the simulation. By utilizing optimization in terms of signaling efficiency, we can make the system itself adjust to the perturbation.
In this way, we are studying signal transduction as a self-organizing system, which uses objective functions to adapt. This basic idea is extremely powerful, and could be used with different kinds of constraints on parameters, reaction times, etc. and with different, multiple input scenarios for larger systems. To explore this question further is of significant importance in assessing cellular health and functioning.

\section{Methods}

\subsection{Example System: GPCR signaling in a submembrane compartment}
Fig.~\ref{fig1a}A shows the example system, a submembrane compartment with a GPCR (G-protein coupled receptor) pathway from a mouse embryonic fibroblast, with ISO as input (extracellular ligand to $\beta(2)$-adrenergic receptors) and the phosphorylation of a protein VASP as output. This model was implemented as an ODE model with 23 reactions and  27 molecular species, derived from 12 initial concentrations 
(cf. Table~\ref{table-rates}, Table~\ref{concentrations}). The parameters were adapted to experimental biological data (not shown, \cite{Blackman2011}). 
In this subsystem, the central cAMP response often follows a plateau curve, i.e. a rise to steady-state, but cAMP transients which are typically observed in cytoplasm may also occur \cite{Blackman2011}.
The dose-response transfer functions were derived as in \cite{Scheler2012}. 
\begin{table}[htb]
\noindent
\begin{center}
\footnotesize
\begin{tabular}{llll}
& $k_{\mbox{\scriptsize\em on}}$&$k_{\mbox{\scriptsize\em off}}$&$k_{\mbox{\scriptsize\em cat}}$\\
 $ b2 + L \leftrightarrow b2L$ & 0.0003 & 0.1 & \\
 $ b2L + PKAc \leftrightarrow b2LPKAc \rightarrow pb2L + PKAc$ & 0.00026 & 1 & 5.4\\
 $ pb2L \rightarrow b2L$ & & & 0.1\\
 $ GsaGDP + b2L \leftrightarrow b2LGsaGDP \rightarrow GsaGTP + b2L$ & 0.006 & 0.8 & 0.2\\
 $ GsaGTP + RGS \leftrightarrow RGSGsaGTP \rightarrow GsaGDP + RGS$ & 0.0008 & 1.2 & 16\\
 $ GiGDP + pb2L \leftrightarrow pb2LGiGDP \rightarrow GiGTP + pb2L$ & 1.2 & 0.8 & 16\\
 $ GiGTP + RGS \leftrightarrow RGSGiGTP \rightarrow GiGDP + RGS$ & 1.2 & 0.8 & 16\\
 $ GsaGTP + AC6 \leftrightarrow AC6Gsa$ & 0.00385 & 3 & \\
 $ GiGTP + AC6Gsa \leftrightarrow AC6GsaGi$ & 0.00385 & 10 & \\
 $ AC6Gsa + PKAc \leftrightarrow AC6Gsa\_PKAc \rightarrow pAC6Gsa + PKAc$ & 0.00026 & 1.5 & 30.4\\
 $ pAC6Gsa + PP1 \leftrightarrow pAC6PP1 \rightarrow AC6Gsa + PP1$ & 0.0026 & 3 & 54\\
 $ ATP + AC6Gsa\leftrightarrow AC6Gsa\_ATP \rightarrow cAMP + AC6Gsa$ & 6e-05 & 10 & 80.42\\
 $ ATP + pAC6Gsa \leftrightarrow pAC6Gsa\_ATP \rightarrow cAMP + pAC6Gsa$ & 6e-05 & 10 & 8.042\\
 $ ATP + AC6 \leftrightarrow AC6\_ATP \rightarrow cAMP + AC6$ & 0.0001 & 120 & 0.142\\
 $ cAMP + PDE4B \leftrightarrow PDE4BcAMP \rightarrow AMP + PDE4B$ & 0.03 & 77.44 & 19.36\\
 $ 1 PKA + 2 cAMP \leftrightarrow 1 PKAr2c2cAMP2$ & 3.5e-08 & 0.06 & \\
 $ 1 PKAr2c2cAMP2 + 2 cAMP \leftrightarrow 1 PKAr2c2cAMP4$ & 2.7e-07 & 0.28 & \\
 $ 1 PKAr2cAMP4 + 2 PKAc \leftrightarrow 1 PKAr2c2cAMP4$ & 8.5e-08 & 0.05 & \\
 $ VASP + PKAc \leftrightarrow VASPPKAc \rightarrow pVASP + PKAc$ & 0.00026 & 1.5 & 30.4\\
 $ pVASP + PP1 \leftrightarrow pVASPPP1 \rightarrow VASP + PP1$ & 0.0026 & 3 & 54\\
 $ AMP \rightarrow ATP$ & & & 1\\
 $ ATP + AC6GsaGi \leftrightarrow AC6GsaGi\_ATP \rightarrow cAMP + AC6GsaGi$ & 6e-05 & 10 & 1\\
 $ b2L + bARR \leftrightarrow b2LbARR$ & 0.0006 & 0.1 & \\
\end{tabular}
\end{center}
\caption{Kinetic rates of the sample system, adapted 
from Sabio-Rk \cite{Sabio-Rk} and Brenda \cite{BRENDA}, 
adjusted to experimental cAMP time-series data \cite{Blackman2011}}
\label{table-rates}
\end{table}

\begin{table}[htb]
\noindent
\begin{center}
\begin{tabular}{llll}
$PDE4B$ & 200&
$b2$ & 100\\
$bARR$ & 500&
$PKA$ & 500\\
$cAMP$ & 100&
$AC6$ & 1000\\
$GsaGDP$ & 200&
$GiGDP$ & 200\\
$RGS$ & 100&
$VASP$ & 200\\
$PP1$ & 100&
$ATP$ & 1e+06\\
\end{tabular}
\end{center}
\caption{Initial concentrations (in nM) of species in the sample system}
\label{concentrations}
\end{table}

\subsection{Objective Function}
\label{methods2}
A biological signal transduction system is defined by its state variable vector $x$, the set of all kinetic rate and initial concentration parameters.

We hypothesize that an efficient signal transmission would maximize the response 
coefficient $R_{C,S}$ (the response of species $C$ to input $S$) defined as
\[ R_{C,S}= {{C_{t}\over C_{0}}-1 \over {S_{t}\over S_{0}}-1}\]
%\[ R_{C,S}^{log}= {{C_{t}\over C_{0}}-1 \over log({S_{t}\over S_{0}})}\]
with concentration change of target $C$ and input $S$
from baseline (t=0) to signal time t. For $R_{C,S}$, values $<1$ show signal loss, with 1 for perfect transmission, and values $>1$ showing signal amplification. 
We may also optimize for the slope $s$ of the sigmoid at half-maximum concentration.  This is equivalent to maximizing $R_{C,S}$, provided that the input signal remains entirely between the upper and lower boundaries of the sigmoid 
(Fig.~\ref{fig1a}). By optimizing for $s$, additional to $R_{C,S}$, we may force the system to implement a switch-like function instead of a more linear function. However, shifting the sigmoid function to the left or to the right is more important as the slope in our models.

In addition, we optimize for reaction time (delay to steady-state). Steady-state is defined, pragmatically, as relative change of less than 2\% over 100s. The delay ($d_S$)
is computed for 90\% (EC90) of steady-state. 
We may now define an aggregate objective function:
\[ f(x)= max_x [R_{C,S}(x), -d_S(x)] \]
%\[ f(x)= max_x [R_{C,S}(o)/ R_{C,S}(x)+ d_S(x)/ d_S(o)] \]
to select the system state variables that minimize delay and maximize response.

In addition to signal transmission from extracellular concentration changes onto steady-state concentrations, such as they typically occur for temporally integrating proteins like transcription factors, %pCREB
we also look at signal transmission by transients, i.e. peak concentration, in response to extracellular signals. %pVASp, pLCa.
In this case, we minimize the delay to peak value, maximize the response at peak value, and minimize the response at steady-state value.
\[ f_{transient}(x)= max_x [R_{C,S}^p(x),  -d_{S}^p(x), -R_{C,S}(x), d_{S}(x)] \]

\section{Results}
\subsection{Delay vs. Efficiency Trade-off}
\begin{figure}[tbp]
\begin{minipage}[ht]{0.35\textwidth}
\begin{tabular}{c}
 \includegraphics[width=2.5in,height=1.8in]{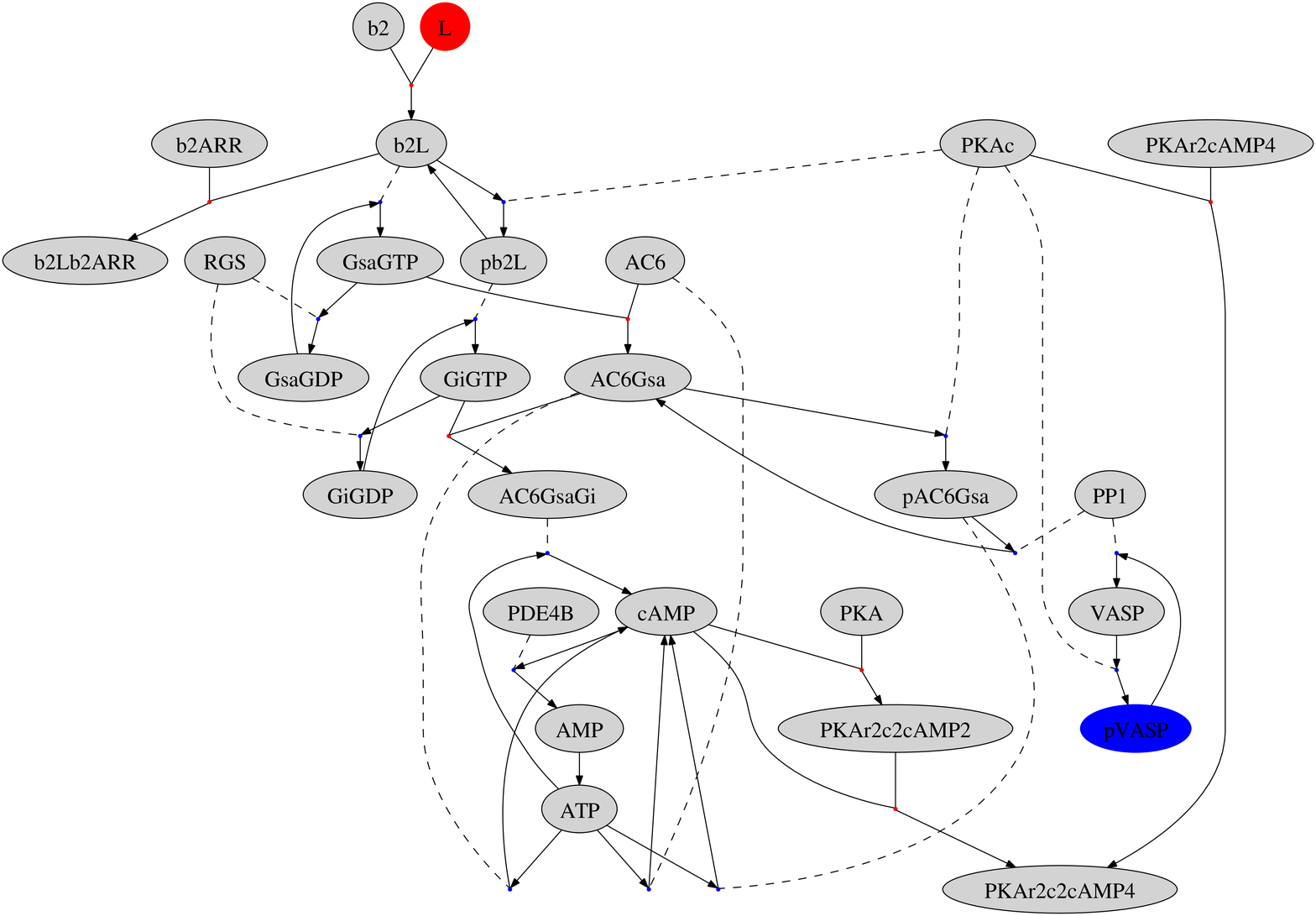}
   \end{tabular}
\end{minipage}\ \
 %\includegraphics[width=2in,height=1.4in]{L_VASP_Blackman_small.tif}
%\hspace*{-0.3cm}
\begin{minipage}[ht]{0.3\textwidth}
\begin{tabular}{c}
  \includegraphics[width=1.7in]{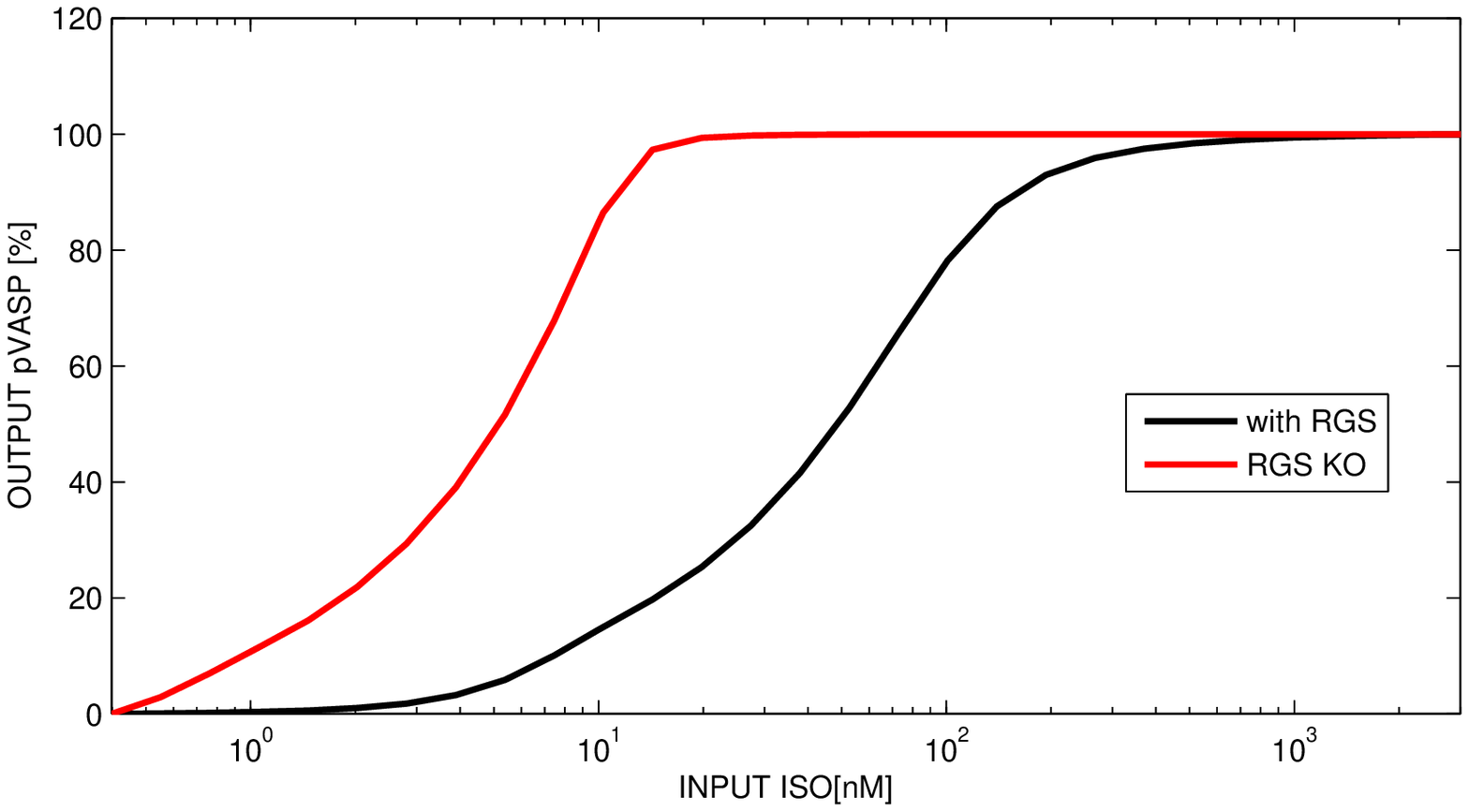}\\
  \includegraphics[width=1.5in,height=1in]{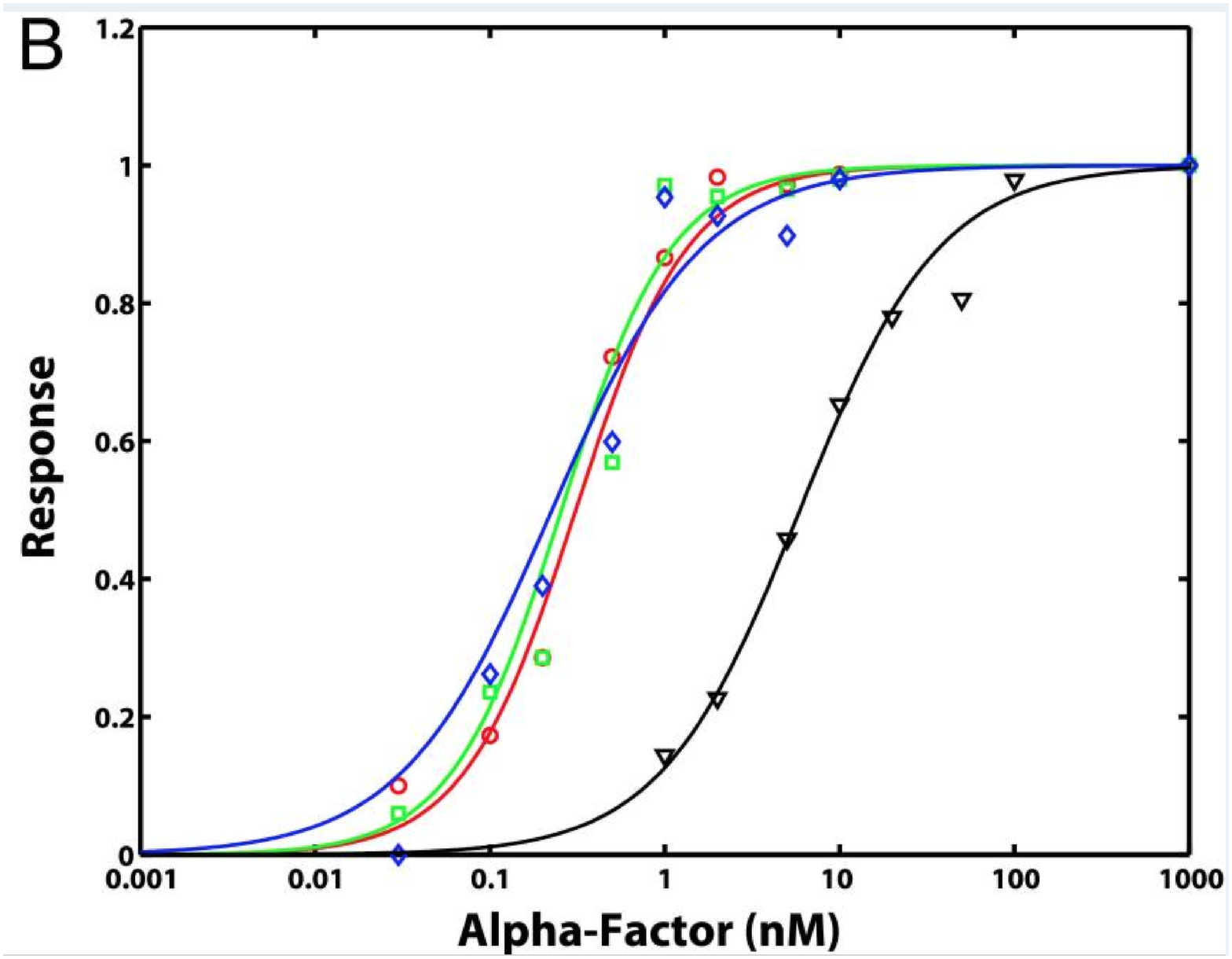}
  \end{tabular}
\end{minipage}\ \ 
%\hspace*{-0.3cm}
\begin{minipage}[ht]{0.3\textwidth}
\begin{tabular}{c} 
 \includegraphics[width=1.5in]{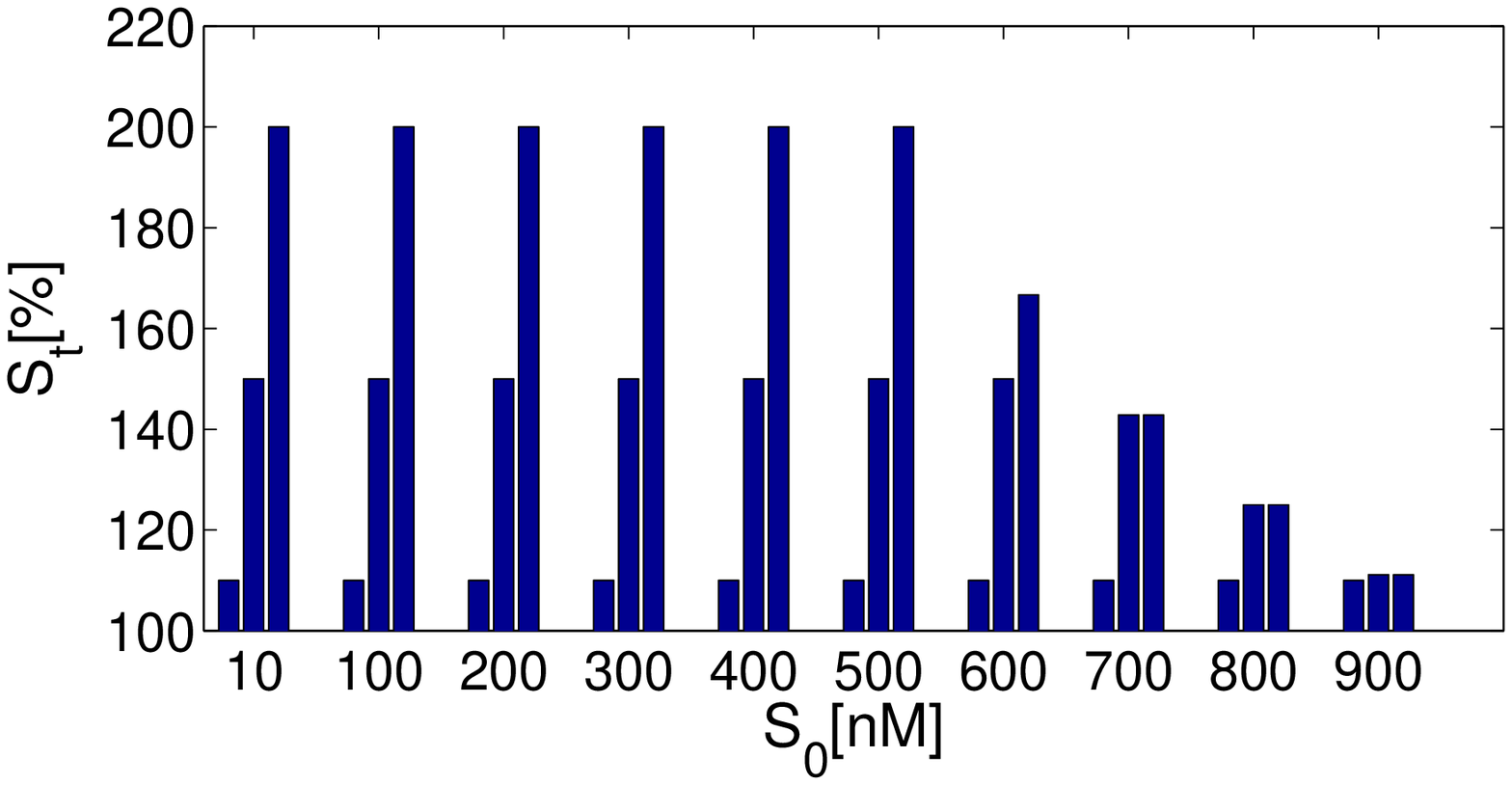}\\
    \includegraphics[width=1.5in]{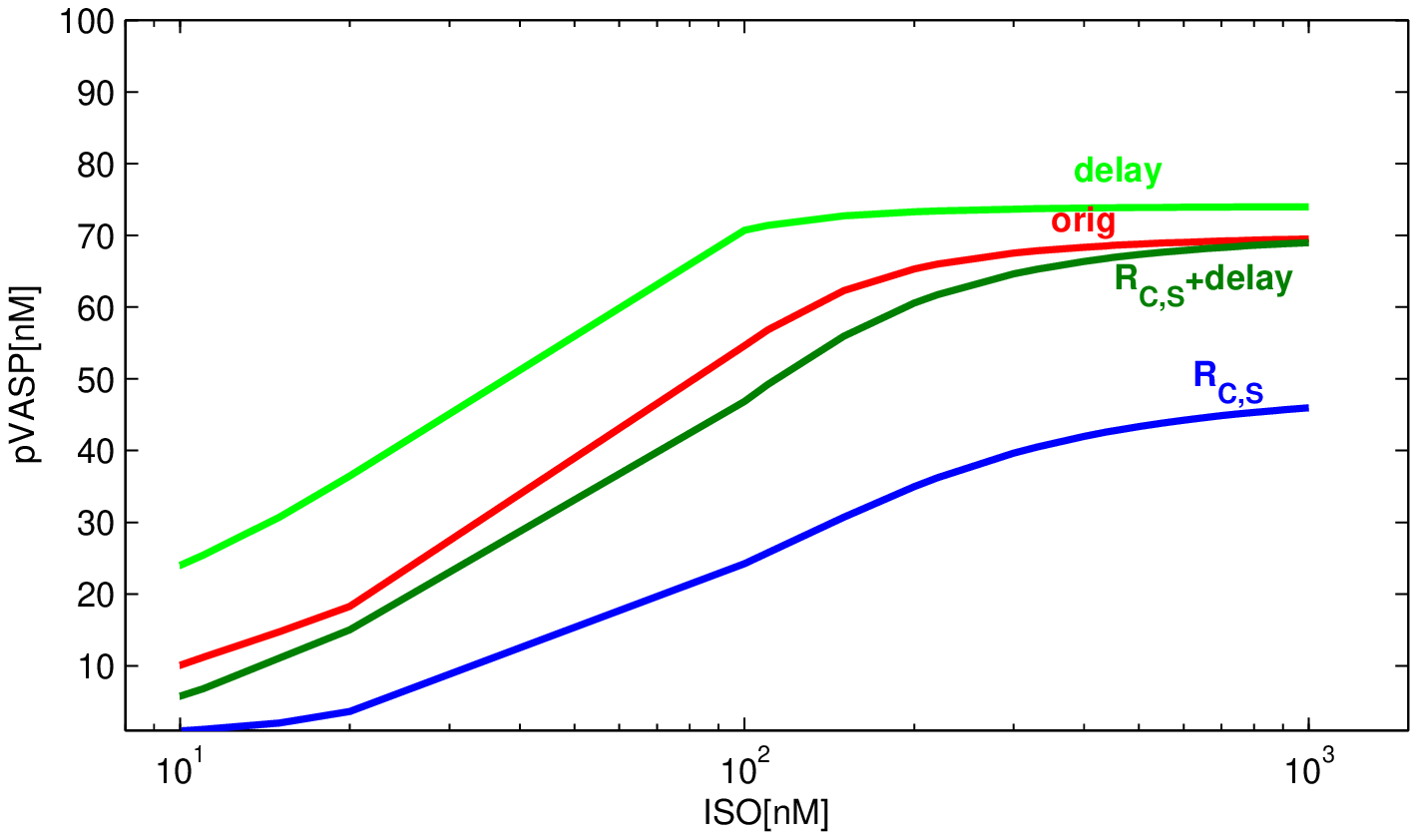}
  \end{tabular}
\end{minipage} 
\caption{A. Biochemical Reaction System with Selected Input(red) and Output(blue) B. %ISO-pVASP transfer function in the MEF model \cite{Blackman2011} B. 
(Top)~ISO-pVASP transfer function (with RGS KO,red)) in the MEF model \cite{Blackman2011}
(Bottom)~Experimental Response to RGS KO for GPCR signal-response in a yeast model \cite{Yi2003} C.~(Top)~Distribution of Extracellular Signals Used to Calculate Signal Transmission. From a baseline signal $S_0$ [10nM-900nM] the extracellular signaling level 
$S_t$ rises to 110\%,150\% and 200\%, but not above 1$\mu M$. (Bottom)~Transfer Functions for 
ISO $\rightarrow$ pVASP. Shown is the original model, optimization by delay, by $R_{C,S}$, and by both. }
\label{fig1a}
\end{figure}

The computation of signal transmission efficiency will be explained here for a single input-output pathway of cAMP/PKA-mediated transmission in a cellular membrane compartment. It is clear that a complex signaling system may have several inputs, and a large number of outputs or target proteins, and this is especially the case for cAMP-mediated signaling. Nonetheless we will focus on the simple case here to explain the basic principle.
The input is a membrane receptor, a G-protein coupled receptor (GPCR), $\beta(2)AR$,
which is activated by an extracellular pharmacological agonist ISO (isoproterenol); the output is a membrane protein, VASP, which promotes actin filament elongation. Activation 
of the receptor selectively inhibits VASP by phosphorylating VASP to pVASP via protein kinase A (PKA) activation. 
In Fig.~\ref{fig1a}B the original ISO/pVASP transfer function is shown, which we use here for further optimization.   
The system was trained for a signal distribution between 10nM and 1 $\mu$M ISO adjusting both kinetic rate and concentration parameters (Fig.~\ref{fig1a}A).  
The optimization method used is a simplex algorithm \cite{Press2007}.
Fig.~\ref{fig1a}B shows the transfer function before and after adjustment,
maximizing for $R_{C,S}$, $d_S$ or the combined function $f$. The results are summarized in Table \ref{exp1}. We see that optimizing for the response coefficient alone shifts the function to the right to better cover the input range. Optimizing for the delay alone shifts it to the left, speeding up signals in the lower range (which are slower). Both objectives together (equally weighted) are fairly close to the original, biologically validated curve, with an improved $R_{C,S}$. 

\begin{table}[htb]
\noindent
\begin{center}
\begin{tabular}{lll}
&$R_{C,S}$&$d$\\
original&0.272 & 423\\
optimized for $R_{C,S}$ &0.41&594\\
optimized for $d_S$ &0.11&169\\
optimized for $f$ & 0.34&481\\
\end{tabular}
\end{center}
\caption{Reaction times ($d_S$) and response efficiency ($R_{C,S}$) in a biological system and under optimization}
\label{exp1}
\end{table}

In general, biochemical reactions are faster at higher substrate concentrations, but the relative concentration change in response to an increase in the enzyme or the binding partner is less. This is a fundamental trade-off between delay and transmission efficiency that may define an optimal operating range for a signaling system and be of relevance in disease processes \cite{Scheler2012}.  The results obtained with this experiment are simple, intuitive and encourage continuing to explore the basic idea.

\subsection{Evolutionary Learning of Kinetic Rate Parameters}
%\vspace*{-0.2cm} 

In principle, we may use all parameters in a system, concentrations or kinetic rates, to maximize signal transmission. But the evolution of protein structure and interactions shows that it is fine-tuning of molecular kinetic parameters which is subject to evolutionary learning, while concentrations are often regulated adaptively in each cell. Mass-action kinetics approximate molecular kinetic parameters, even though there are  significant sources of uncertainty, such as the stochastics of molecular interactions.
In the following, we explore the idea that kinetic rate learning operates on evolutionary time-scales, and that biologically attested signal transduction pathways contain reaction rates which are optimal in terms of signal transmission efficiency.
We use known concentration ranges, specific by cell type, together with kinetic rate optimization. 

\begin{figure}[tbp] % float placement: (h)ere, page (t)op, page (b)ottom, other (p)age
\begin{minipage}[ht]{0.35\textwidth}
 \begin{tabular}{c}
   \includegraphics[width=2in,height=0.5in]{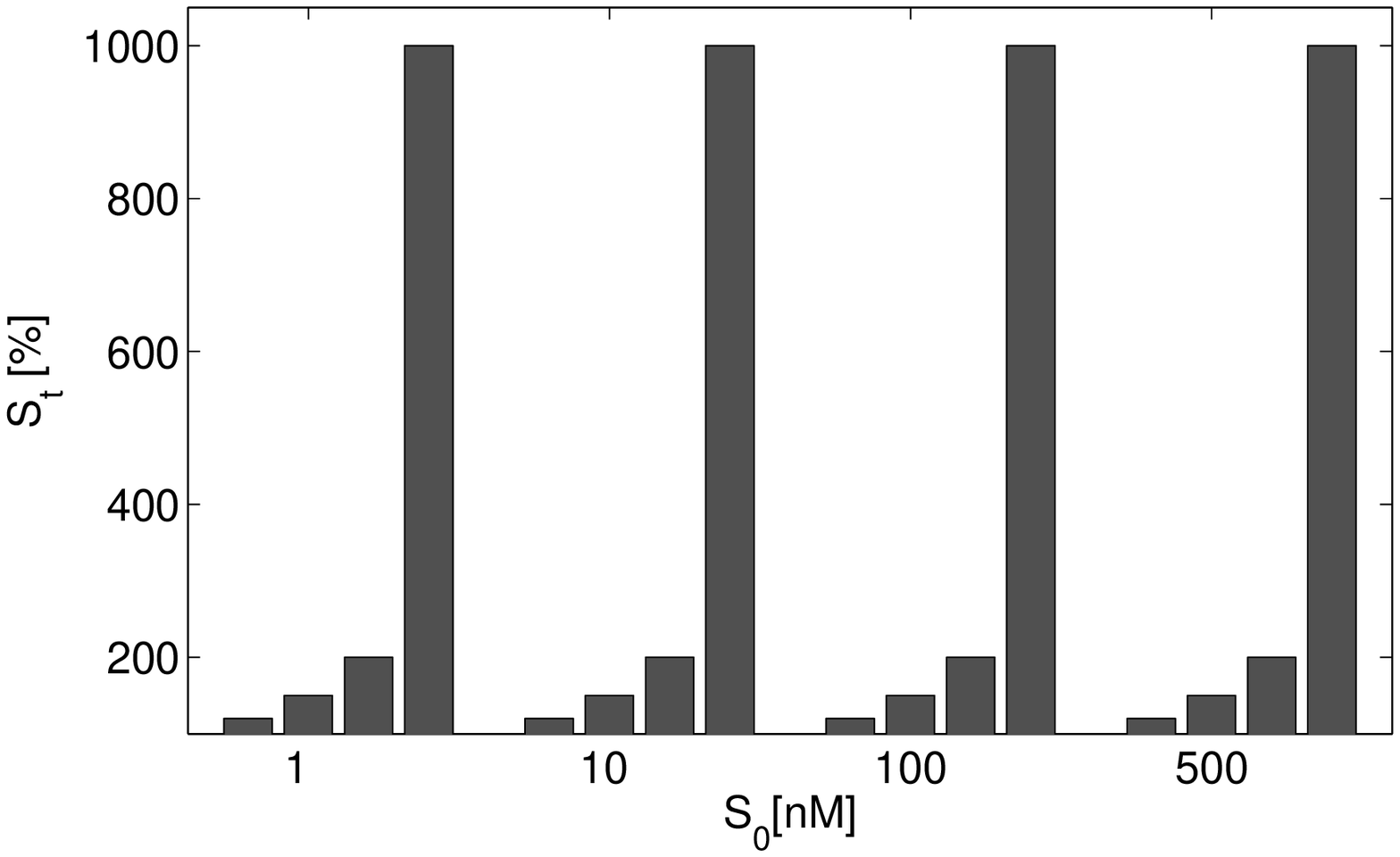}\\
     \includegraphics[width=2in]{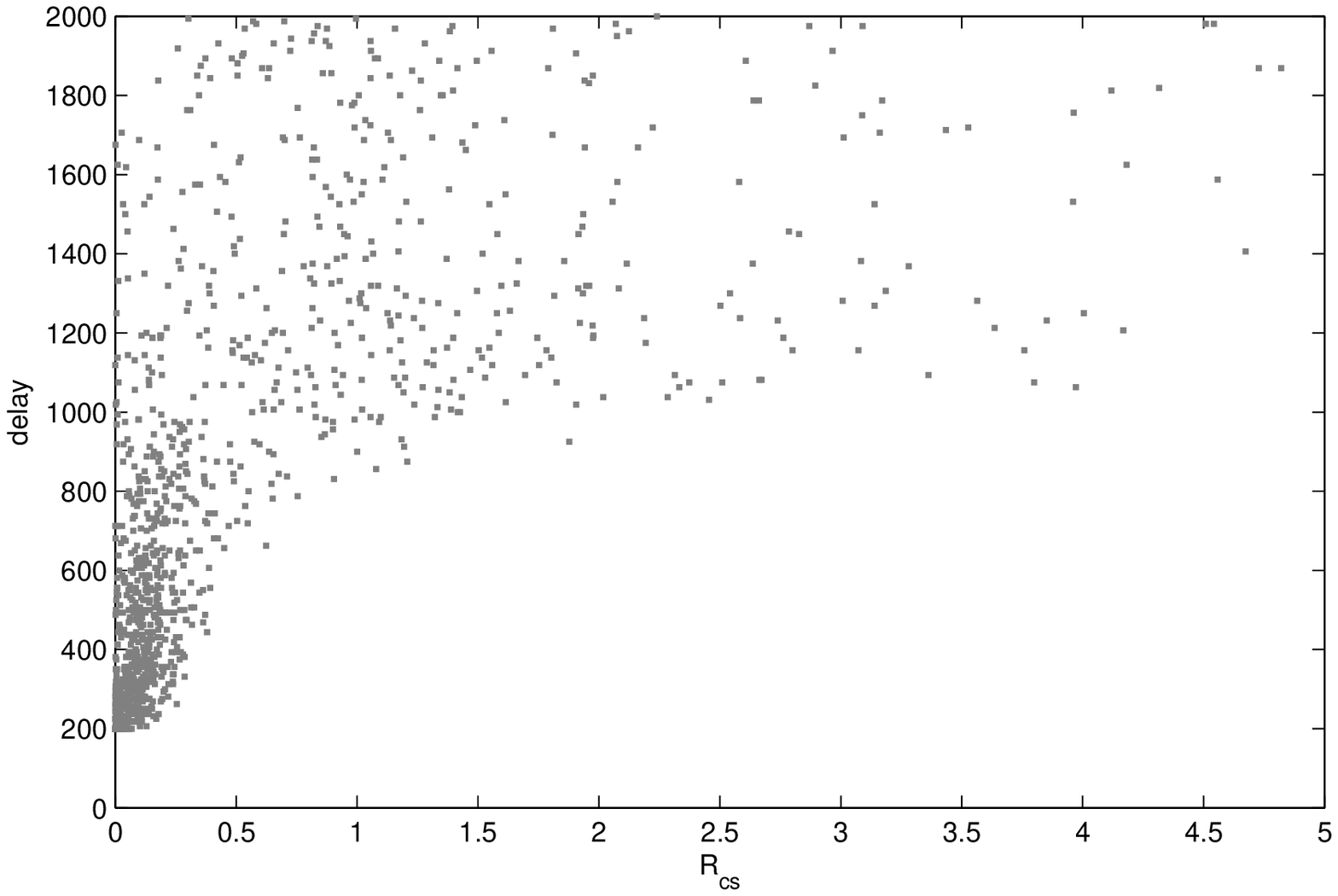}
  \end{tabular}
\end{minipage}\begin{minipage}[ht]{0.3\textwidth}
  \includegraphics[width=2.5in,height=2in]{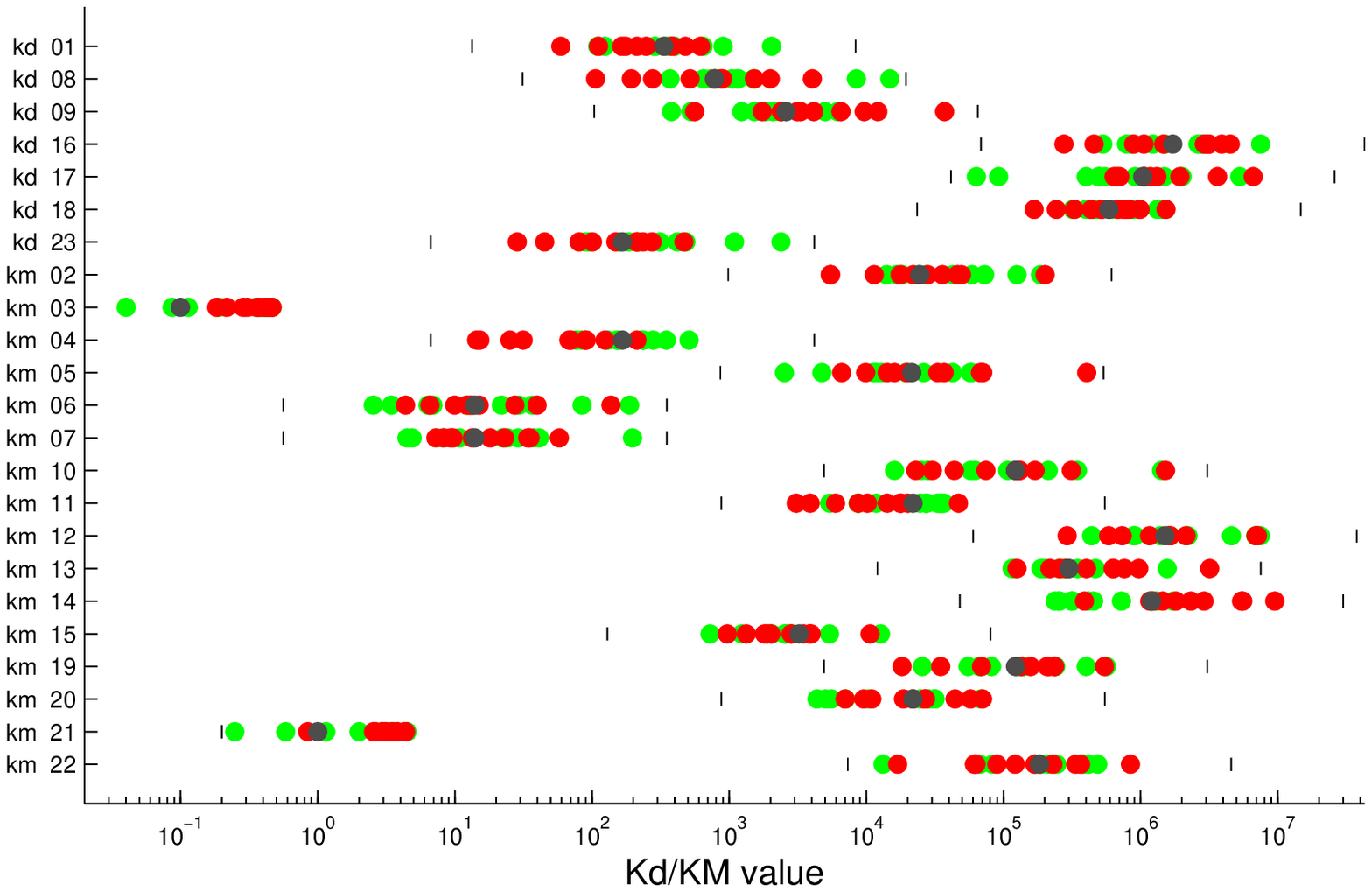}\end{minipage}
\begin{minipage}[ht]{0.3\textwidth}
 \includegraphics[width=2.2in,height=1.8in]{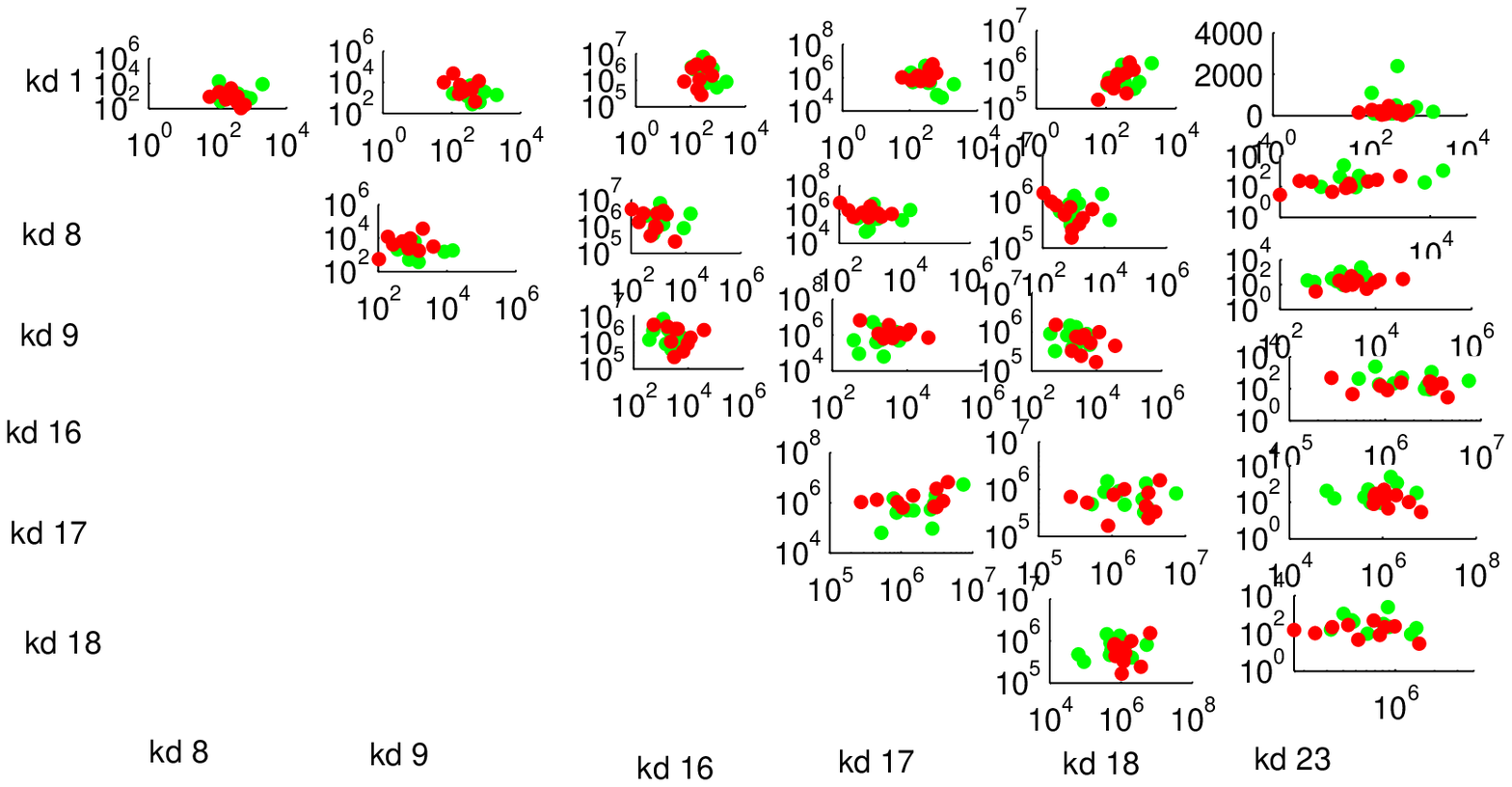}\vfill
\end{minipage}
  \caption{A. Distribution of systems according to R and d values. $\sim$ 1600 different systems states were randomly generated with k-parameter variations (20-500\%), $R_{C,S}$ and $d$ values were measured with the signal set shown on top. B.~Kd and KM Parameters for high efficiency (red, $f<2$) and low efficiency (green, $f>5$) systems. 
C.~Cross-correlation of Kd values.}
  \label{fig:param}
\end{figure}

To explore the parameter space, we drew 1662 values for all kinetic rate parameters (kon, koff, kcat) from a distribution of 20\% to 500\% of the original parameters, and calculated 
$R_{C,S}$ and $d_S$ for the corresponding models, relative to an improved signal distribution from $1nM$ to $1\mu M$ (Fig.~\ref{fig:param}A). We find that there are parameter combinations which greatly improve efficiency of the signal transmission function. The basic distribution of a uniform low value of $R_{C,S}$ for fast $d$ and a wide variability in $R_{C,S}$ above a certain threshold in $d$ is robust against different types of signaling input (cf.~also Fig.~\ref{transient}A). This may, however, be highly dependent on the reaction network that underlies the transfer function. We have not further explored this question.
To test for robustness of these systems against variability of concentration, we repeated experiments for 100 systems with 20\% variation of original 
concentrations, which corresponds to generally accepted noise levels 
(cf.~\cite{Sigal2006,Newman2006}). As expected, this low variation did not significantly affect the quality of a set of kinetic rates (supplemental table).

We further analysed the parameter combinations with different signal transmission efficiency. In Fig.~\ref{fig:param}B and C, we distinguish low and high efficiency signal transmission.
Interestingly, we find, with respect to signal efficiency of the transfer function, that all parameters are 'sloppy' (allow a wide variation), there are no 'stiff' (low variation) parameters \cite{Gutenkunst2007}. Standard deviation for all parameters in our case is similar (Fig.~\ref{fig:param}B). Since it has been argued that optimization to experimental data yields reactions which allow more variance than others, as an indication of their influence on the signal transmission pathway, this analysis seems to contradict this effect. 
Possibly, these results pertain mostly to parameter variation that results from matching a networked system with many species to selected time-series data for only few species, which may behave differently from general optimization.

\subsection{Co-regulation of Protein Concentration as an Adaptive Response to Perturbation}
%\vspace*{-0.2cm}
Co-regulation of protein expression in cellular systems is important in disease progression and often a problem in targeted interventions. %drug resistance 
Here we are exploring the question of self-organization of protein concentration after a perturbation that reduces one protein to only 10\% of its previous concentration. Keeping kinetic rates fixed, all concentrations in the system are allowed to adjust until optimality of signal transmission and delay is reinstated.  There is a number of interesting observations here (cf.~Fig.~\ref{fig:fig_conc_reduct}A), which relate to the biological reality. For instance, reducing PDE4B causes much regulation in other proteins, but it is almost never targeted. In contrast, reducing PP1 has little effect, but PP1 is frequently responsive to other proteins. Reducing PKA, RGS and AC6 leads to widespread down-regulations, to maintain sensitivity of signaling, but reducing the receptor beta-2 leads to up-regulation. There are many individual adaptations which can be interpreted to maintain the sensitivity and responsivity of the small molecule cAMP. The results show that protein regulation is highly sensitive to positions and roles of individual proteins in the reaction network in transmitting the signal. This problem may also be amenable to a more principled mathematical analysis \cite{Steuer2006}.  Another idea would be to rank reaction systems defined by kinetic parameters as in Fig.~\ref{fig:param} by how well they adapt to perturbations.

In our example, the quality of the readjustment is not always the same, but the simple optimization scheme that we use may easily produce suboptimal solutions (local minima). 
Concentration changes may be caused by genetic up- or down- regulation, secretion and re-uptake, increased degradation, RNA interference,  etc. and are therefore not easy to model from the standpoint of mechanistic biological modeling, which would need modules for all biologically attested processes.  A unified perspective by a set of constraints and a set of objectives, such as has been envisaged here, may lead to better predictive results and may also be used as a guiding principle in constructing mechanistic models. Since there are intricate biological processes of adjusting concentrations, we may assume that in the cell optimal solutions are more easily found.

\subsection{Optimal signal transmission depends on the signaling level}
%\vspace*{-0.2cm}

\begin{figure}[tbp] % float placement: (h)ere, page (t)op, page (b)ottom, other (p)age
  \begin{minipage}[ht]{0.35\textwidth}
 \includegraphics[width=2.4in,height=1.6in]{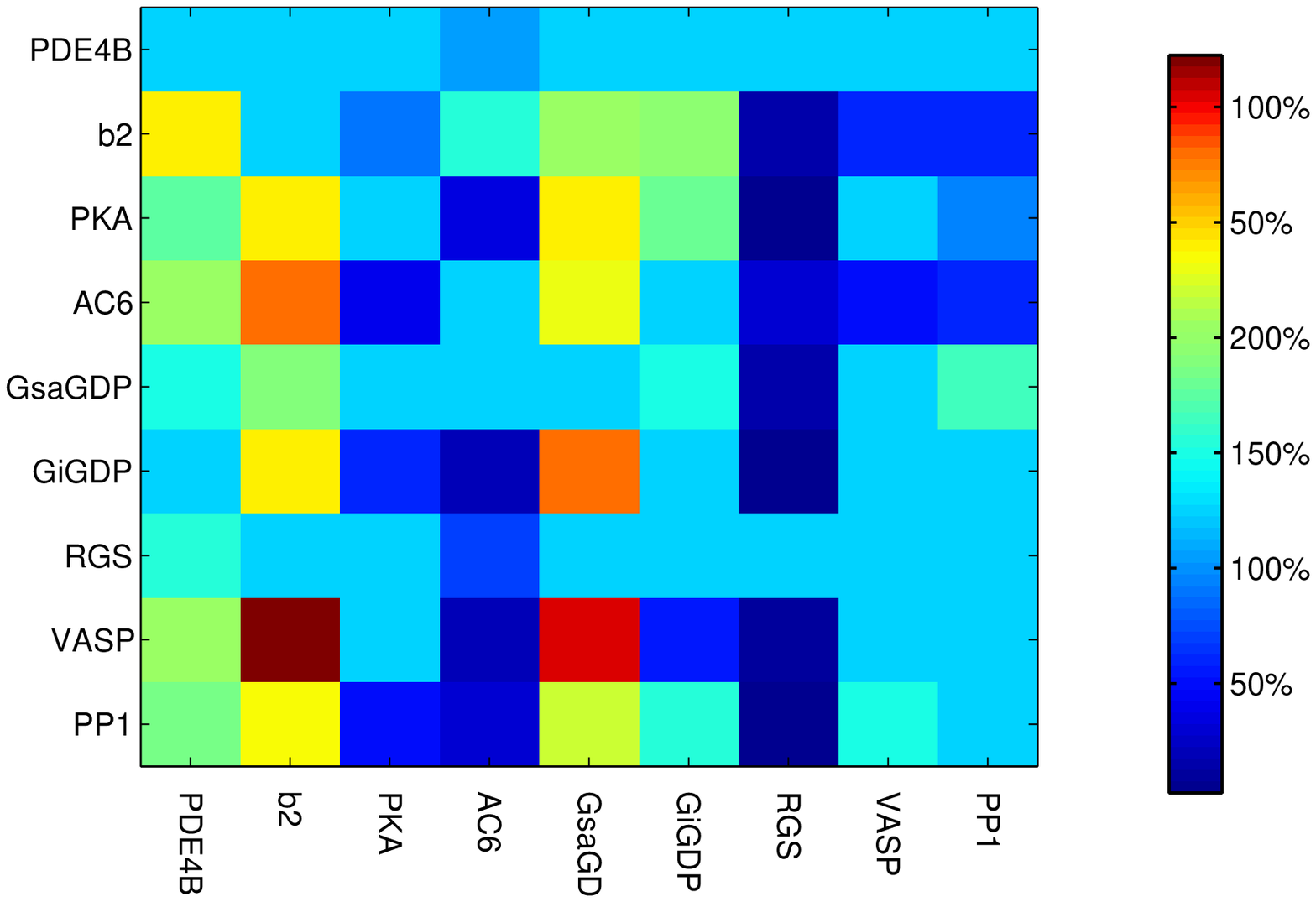}\vfill
\end{minipage} 
\begin{minipage}[ht]{0.3\textwidth}
 \begin{tabular}{c}
   \includegraphics[width=0.8in,height=0.46in]{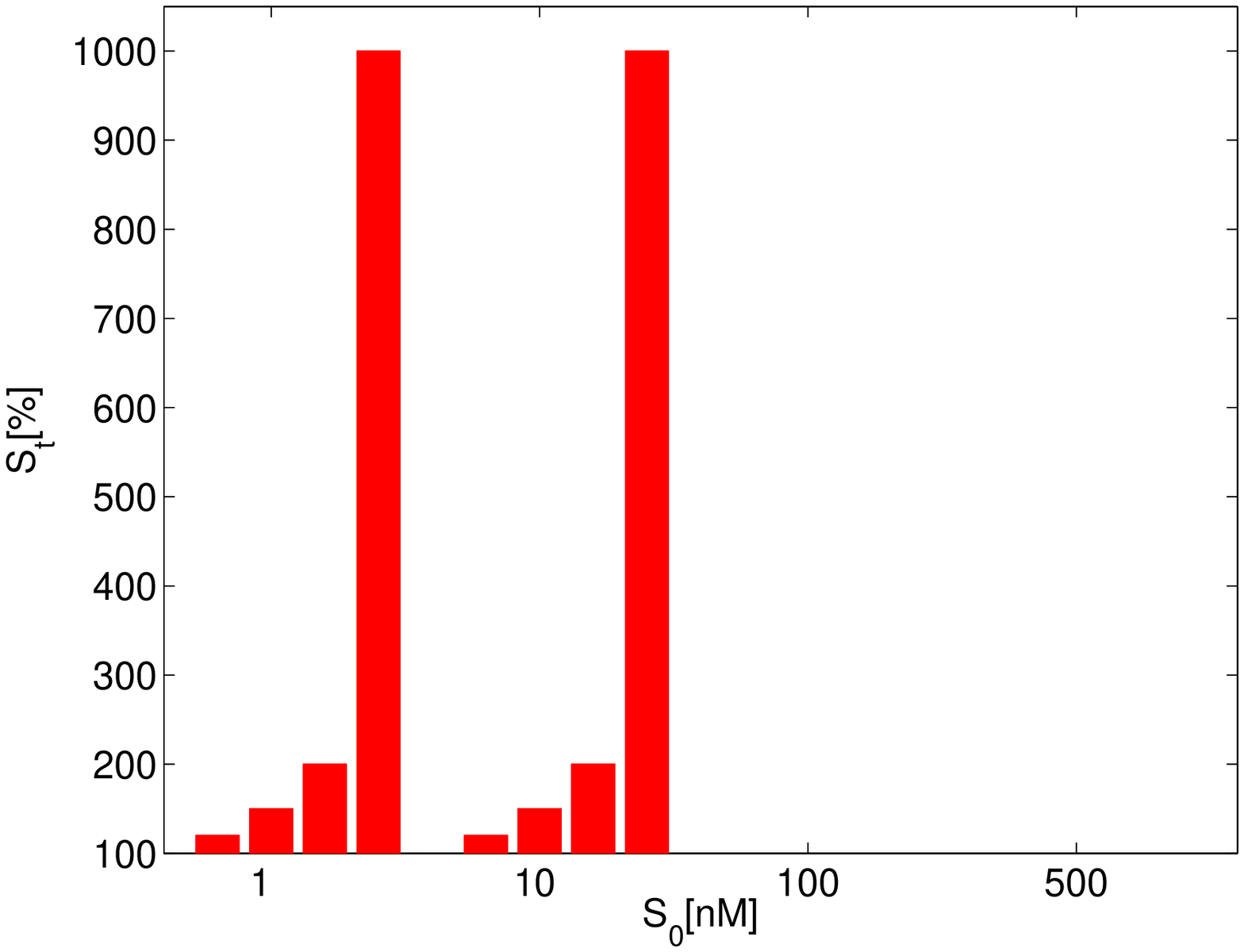}
%R7oc_signals_small_2.eps}
   \includegraphics[width=0.8in,height=0.46in]{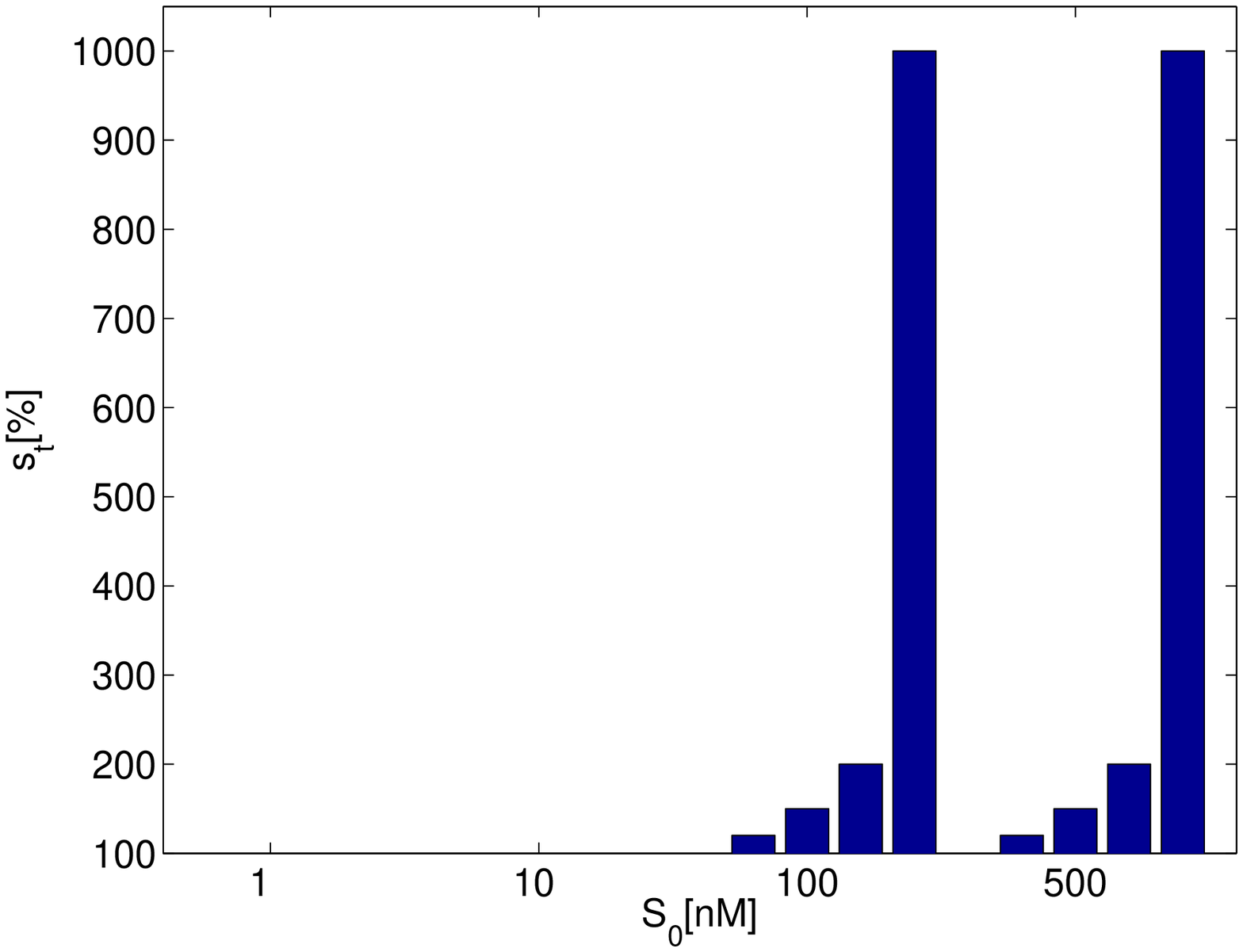}\\%R7oc_signals_large_2.eps}\\
    \includegraphics[width=2in, height=1in]{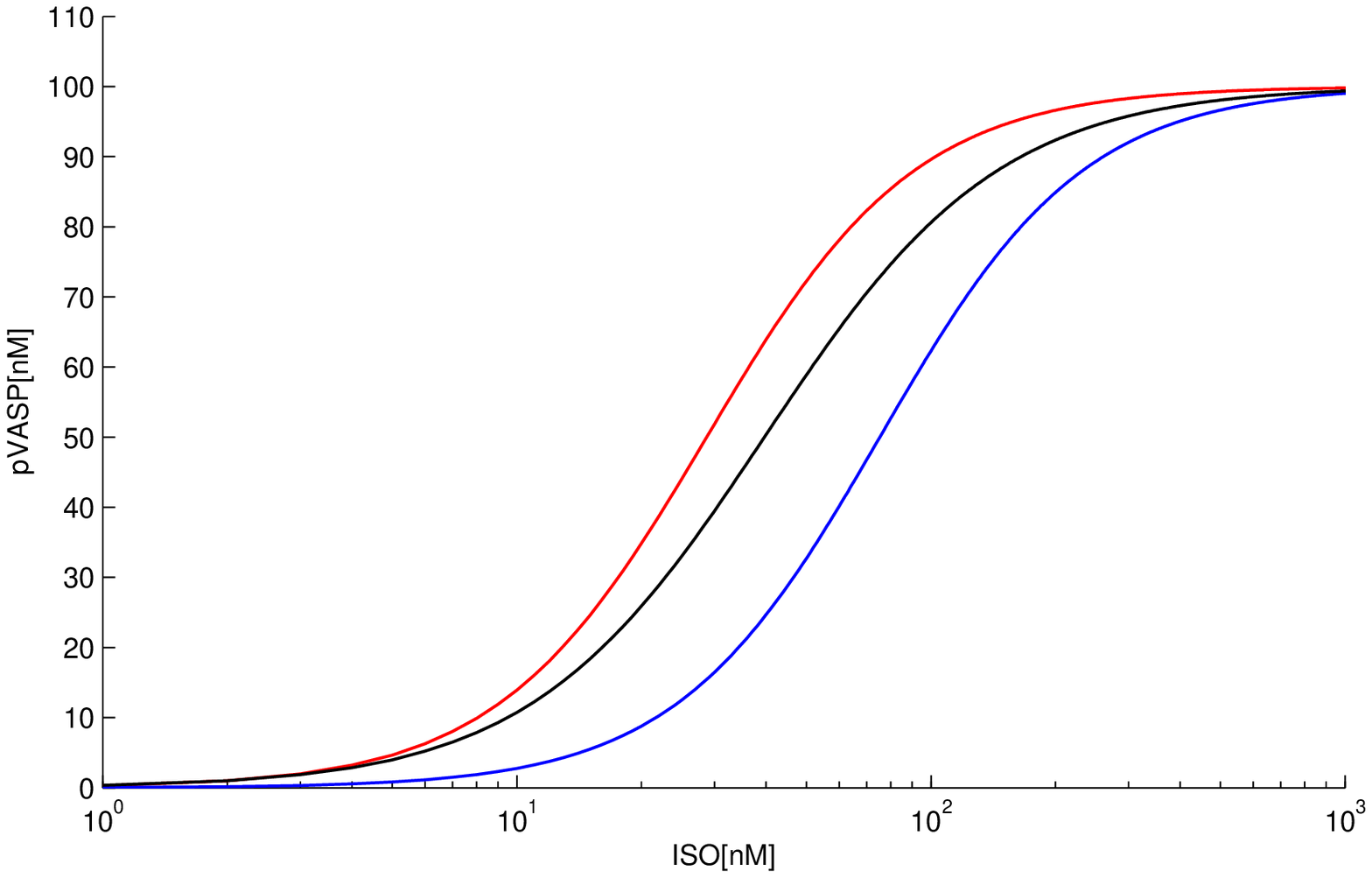}
  \end{tabular}
\end{minipage}\ \  
\begin{minipage}[ht]{0.3\textwidth}
 \includegraphics[width=2.1in,height=1.6in]{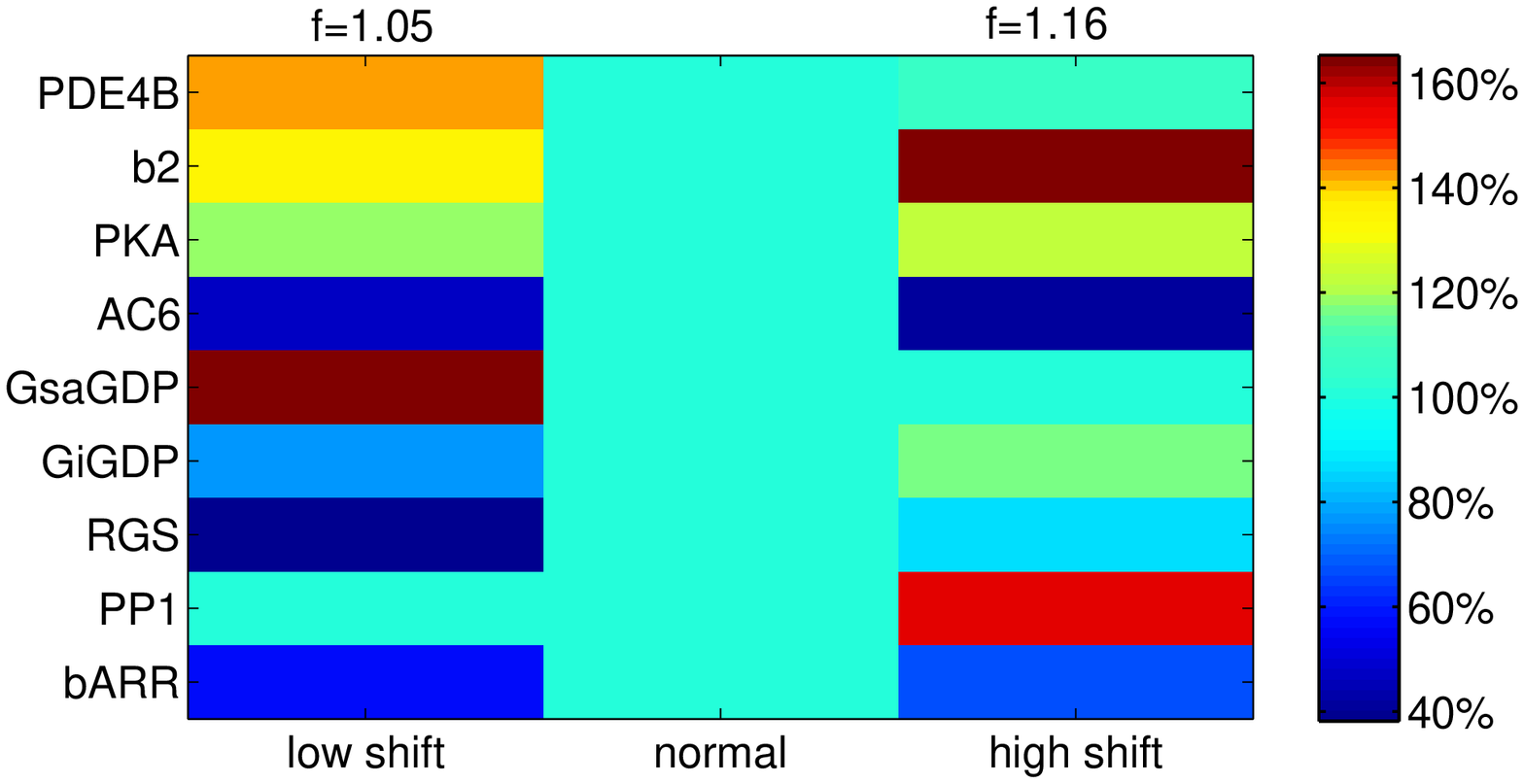}\vfill
\end{minipage} 
  \caption{A. Concentration Changes in Response to Total Protein Concentration Reduction.
Selected concentrations were reduced to 10\% (shown on the x-axis). Learning was applied until $f$ values were improved (shown on top for each experiment). Colors show the relative adjustment of all concentrations on the y-axis. B.~Transfer Function Response to Shifts in Input Signal Range (shown on top).
C.~Concentrations Changes in Response to Extracellular Signal Shift. Matrix is constructed as in A.}
  \label{fig:fig_conc_reduct}
\end{figure}

From the standpoint of disease modeling, an unusual protein concentration may be an adaptive response where a still functioning cell in a dysfunctional external signaling environment struggles to keep signal transmission efficient to support cellular function. In such a case, targeting this protein by pharmacological intervention will lead to co-regulation on other proteins. 
In general, as well as in real biological signaling systems, it is the localization of the input signaling range and the distribution of signals that the system transmits which are important for optimization.
If we select the signal set that we optimize for in the right way, the input range will move towards the loglinear range, i.e. between the lower and upper input boundaries 
(Fig.~\ref{fig:fig_conc_reduct}B). As a result, a number of internal concentrations will change.  
%Fig.~\ref{fig:rgs-yeast} C 
Fig.~\ref{fig:fig_conc_reduct}C shows the relative concentration changes that result from a shift in input range. Interestingly, the low shift requires a strong reduction in RGS, and upregulation for PDE4B, among other effects. GsaGDP (the activating G protein) is strongly increased, and GiGDP (the inhibitory G protein) is decreased. In the high shift, 
GiGDP and PP1 (the phosphatase which decreases pVASP) are stronger, and the beta-2 receptor density is much increased. In the future, it will be interesting not only to calculate these effects based on different optimization measures, and in multiple input-output scenarios, but also to compare this with attested cases of biological adaptation. By reverse engineering, we may infer an optimization measure from a sufficient number of attested co-regulations.
For instance in addiction, protein co-regulation as a form of sensitization is well-attested \cite{Robison2011}, but also in cancer where intercellular signaling (e.g. by cytokines \cite{OShea2008}) is affected. Gene expression data may then be mined not only for evidence of the mechanics of genetic regulatory pathways but also for evidence of shifts in extracellular signaling, which cause altered protein expression.
\subsection{Role of transients}
\begin{figure}[tbp] % float placement: (h)ere, page (t)op, page (b)ottom, other (p)age
 \begin{minipage}[ht]{0.28\textwidth}
 \begin{tabular}{c}
   \includegraphics[width=2in,height=0.8in]{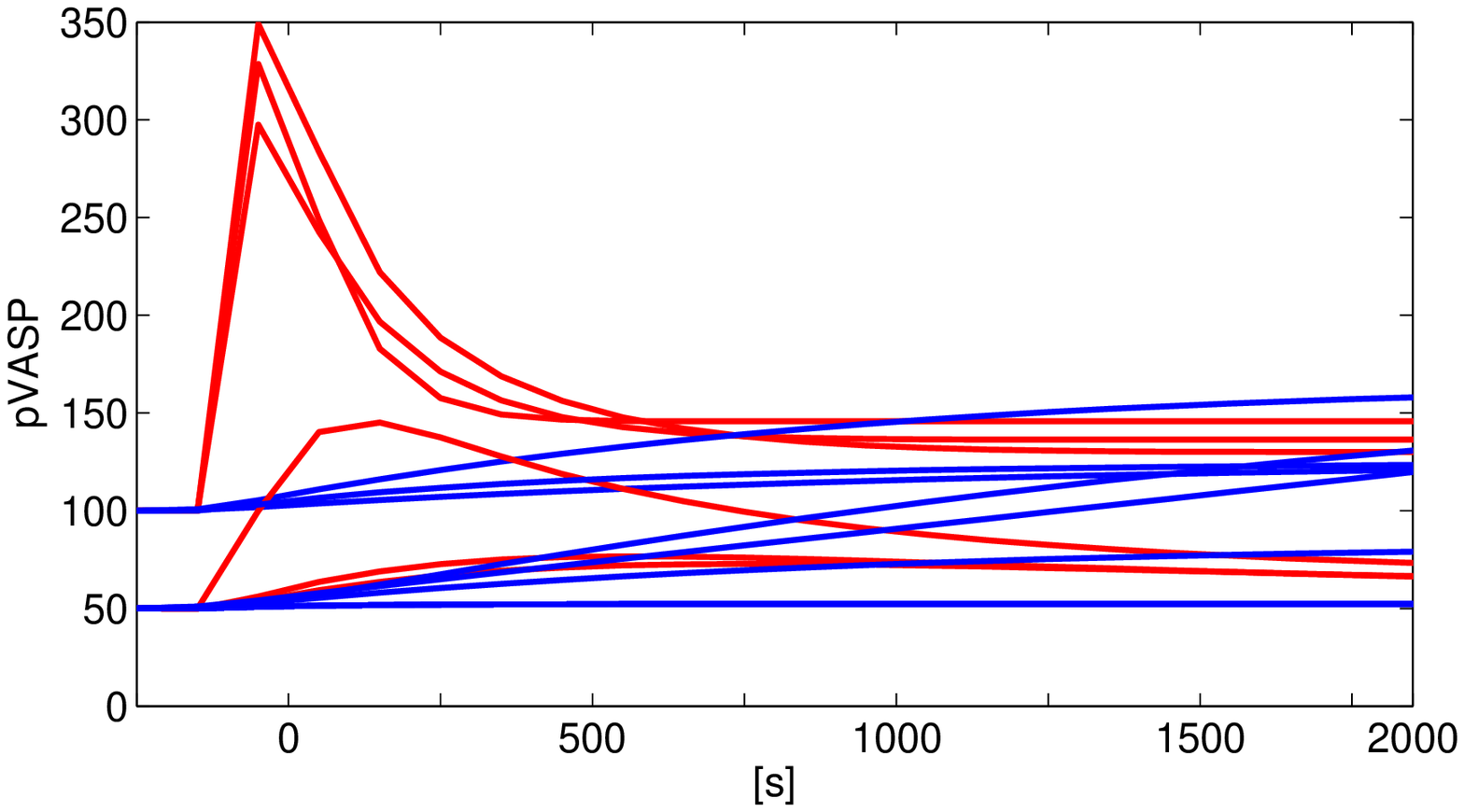}\\
    \includegraphics[width=2in, height=1in]{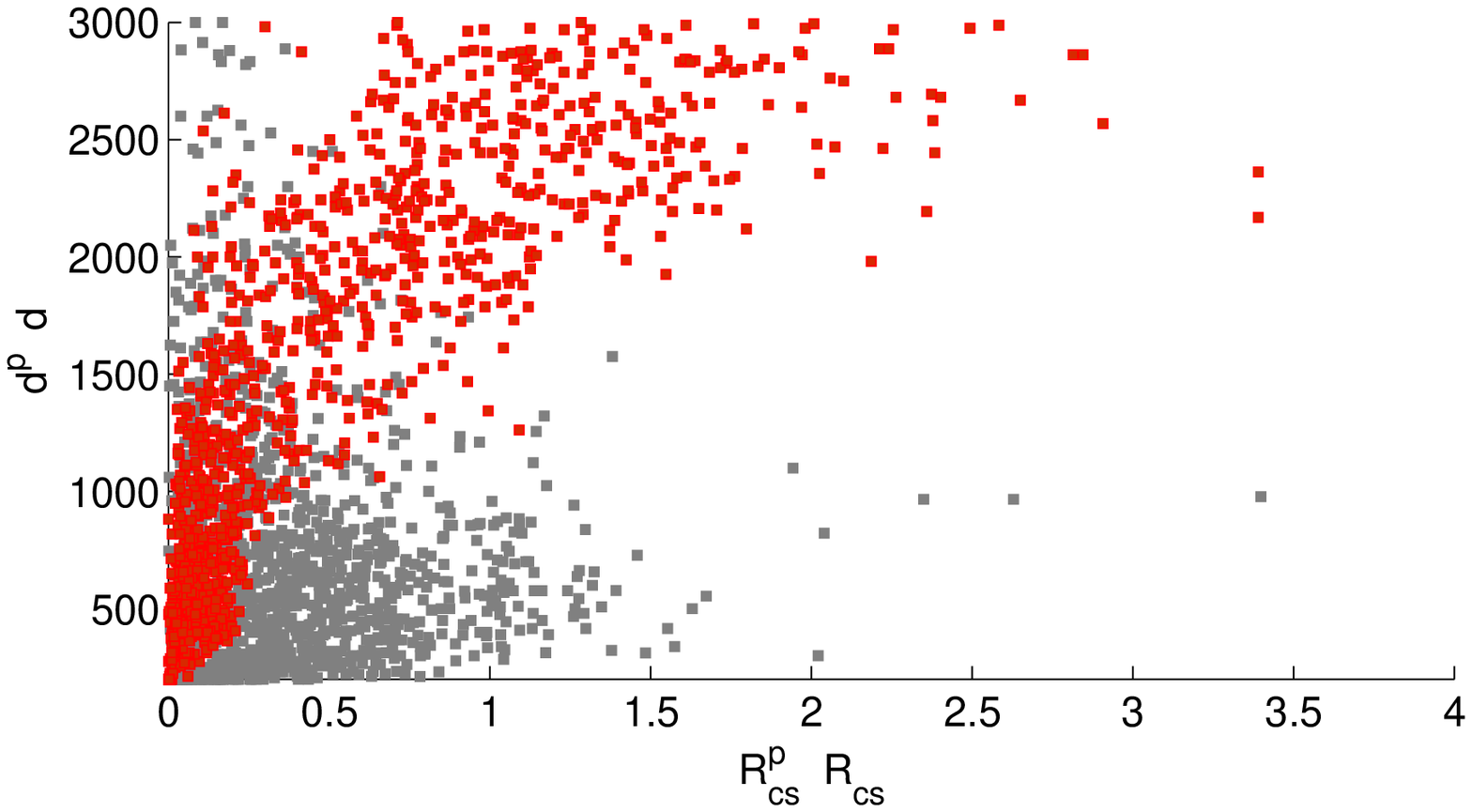}
  \end{tabular}
\end{minipage}\ \  
\begin{minipage}[ht]{0.35\textwidth}
 \includegraphics[width=2.4in,height=1.6in]{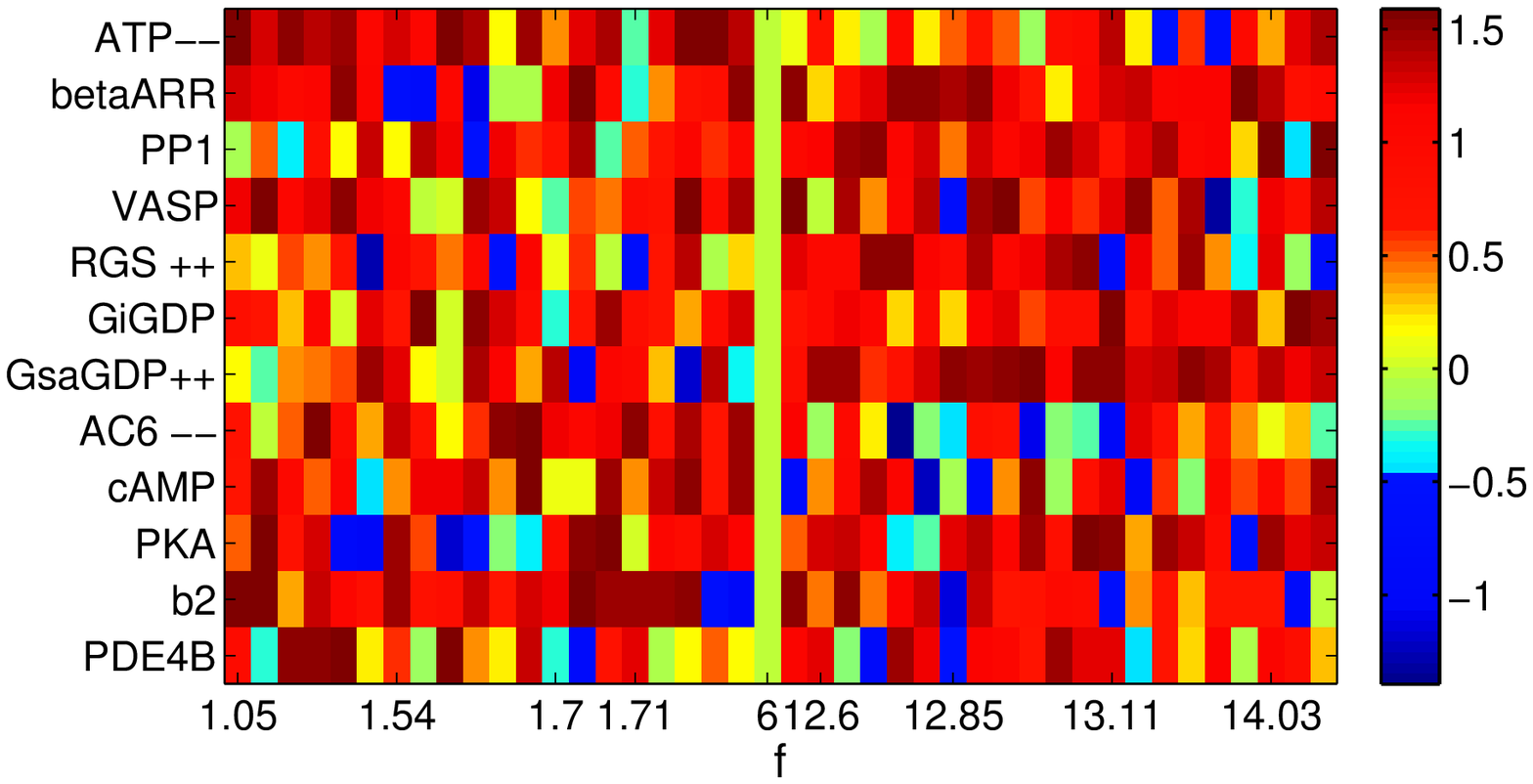}\vfill
\end{minipage}  
\begin{minipage}[ht]{0.35\textwidth}
 \includegraphics[width=2.5in,height=1.8in]{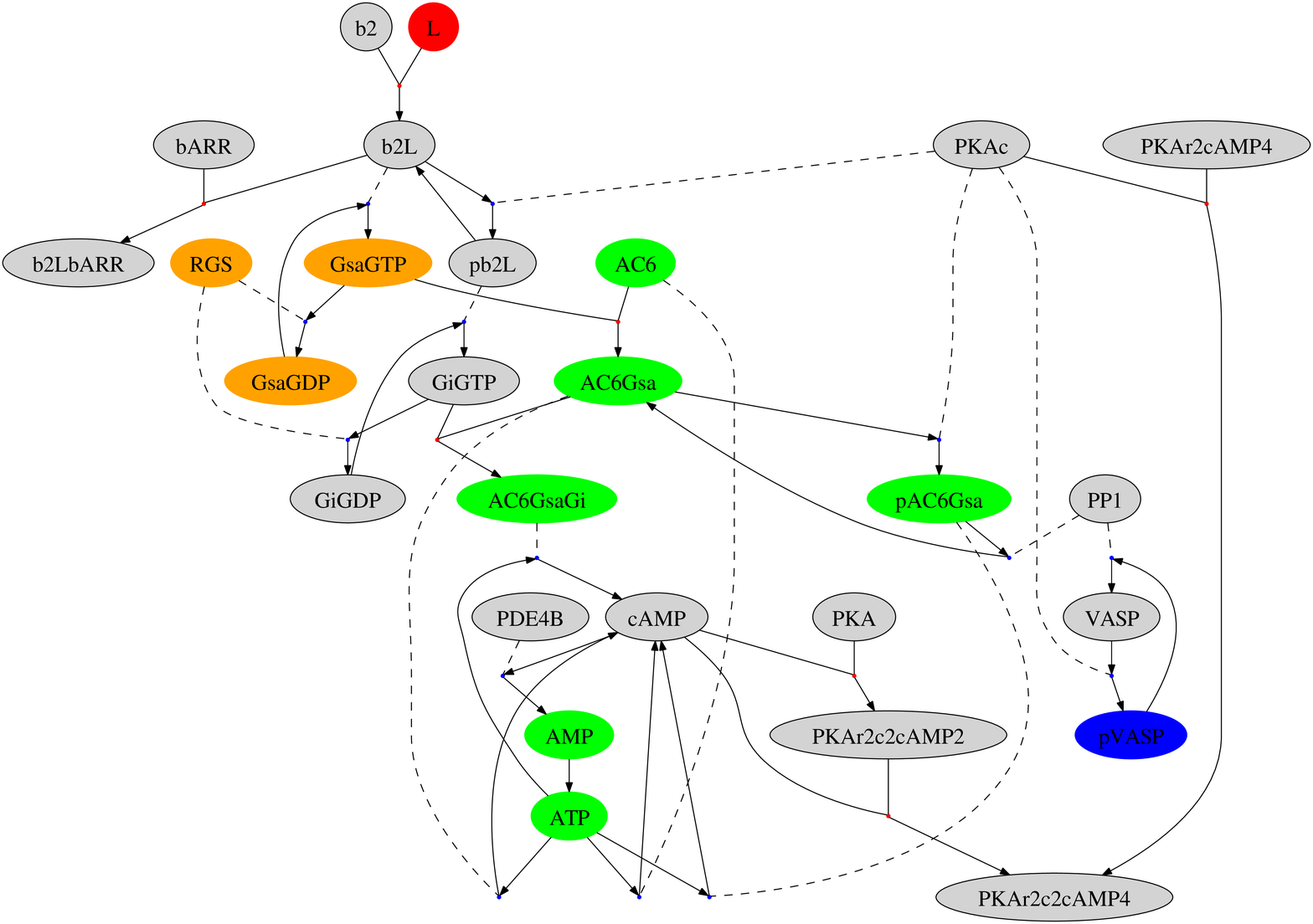}\vfill
\end{minipage} 
  \caption{
 A. (top) Dynamics of selected systems with low ($f_{transient}<3$,blue) or high ($f_{transient}>10$,red) propensity for a transient response.
(bottom) Distribution of concentration parameters according to $d^p$,$R^p_{C,S}$ (grey)
or $R_{C,S}$,$d$ (red). 
B. Concentration changes in response to optimizing for transients, with plateau signaling (left) and strong transients (right), sorted by $f_{transient}$. 
C. Concentration shifts (orange=high, green=low) in the biochemical reaction network for a high transient system, as averaged from B. Conforming to intuition, we see that the earlier, driver complex (G-protein) is high for transients, which increase the on-slope for cAMP production.}
  \label{transient}
\end{figure}

The appearance of a transient vs.~a plateau signal (or even a dampened oscillation) in 
response to sustained signaling depends on the construction of the biochemical reaction network (negative feedback interactions) and its parameters. 
We show that we can also train a system for the appearance of a transient response to 
a sustained signal. This means to search for high response at peak, low delay to peak, but also  a low response at steady-state (cf.~\ref{methods2}, Fig.~\ref{transient}A). This requirement of invariance for the steady-state is sometimes considered a form of 'robustness' or 'homeostatic regulation' of the cellular response
 \cite{Shinar2010,Kitano2007}, and signaling by transients is widely regarded as an important mechanism. 
Here we found that concentration adjustment is quite sufficient to acquire a switch from transient to plateau response, with a high variability of response shape dependent on concentration, but also on the size of the signal (Fig.~\ref{transient}A). In 
Fig.~\ref{transient}B we show the distribution of concentrations that achieve high or low propensity for transients, as indicated by $f_{transient}$. We focus on concentrations with fairly uniform up- or down-regulation to create an experimental graph in 
Fig.~\ref{transient}C. 
The simplicity of Fig.~\ref{transient}C undermines the notion, as proven by the random search optimization, that deviations from up- or down-regulation for individual concentration may not be irrelevant noise, but rather part of a guided adaptation process that has many different solutions. We may have to consider this complexity to understand concentration adjustments in biological cells.

By using different objective functions for each target, it will be quite possible to combine different goals for targets in a multiple-output signaling system. 
Interestingly, a self-organizing cellular signaling system may also incorporate adaptive controls as objective functions, in particular when ion channels or membrane transporters are targeted. In this case, the goal of the system is to respond to an input (e.g. ISO at beta-2 receptors) such that the magnitude of another input such as calcium (ion channels as targets) is controlled. Signal transmission efficiency for such a target is not to be optimized to a maximum value, but instead to the appropriate ratio between the inputs. We have not followed up on this idea, but it may be worth to further investigate as an example of natural computation for adaptive control.

\section{Discussion}
%\vspace*{-0.2cm}
The idea to look for parametric optimization as the basis of realistic cellular properties has been applied with success to metabolic fluxes \cite{Edwards2001}. In that case, optimization of growth is usually regarded as the single objective function. Here we use another objective function, signal transmission efficiency, to study the adaptive response of a signaling system to perturbation in concentrations. This equals maximization of concentration change in a target species in response to input signal. Optimization of growth and optimization of signal transmission are therefore related. 

Signal transmission in a cellular system may have multiple functions: transcription factor activation, which may be related to the cell cycle, to morphological change or to adjustment of protein concentrations, cytosolic kinase/phosphatase activation with multiple cellular targets, membrane protein activation such as ion channels or receptors, etc.
We assume that signal transduction has been optimized by evolution, and that concentration adjustment exists to maintain effective signal transmission. We have shown how to optimize a single-input single-output system for both speed and signal efficiency due to the basic properties of kinetic equations \cite{Scheler2012}.  It is easy to extend the present discussion to optimize for multiple outputs in parallel, and a system could also be optimized for a number of I/O functions. In that case other measures, such as mutual information, may also be employed as objectives. We used optimization in a two-step process: (a) in evolutionary learning, in order to find kinetic rates for estimated concentrations (b) in cellular adaptation, in order to re-calibrate the model in response to perturbations in concentrations, changes in extracellular signaling, or to effect a transient vs.~plateau-like response. This also means that we have transitioned from a biologically defined system that is built bottom-up from available parameters and data to a self-organizing, learning system that adjusts to changes on evolutionary or individual time-scales. 

Matching models to experimental time-series data is an under-constrained problem that often yields large ranges of suitable parameters
 \cite{Secrier2009,Gutenkunst2007}. Analysing signals and targets to find optimal transmission parameters may be used to further constrain and investigate the parametric space. Signal transmission efficiency in a biological system can be measured directly
 \cite{Cheong2011,Brennan2012} 
and be compared to what is theoretically possible given a set of equations. During evolution, new protein subtypes develop with a different set of interactions and regulations. This corresponds to an adjustment of the available set of reactions, a type of structural learning to overcome the bottlenecks that are a result of tightly specified molecular kinetics and limited adjustment of concentrations.

\newpage

\bibliographystyle{plain}
{

}

\newpage

\begin{table}[htb]
\noindent
\begin{center}
\begin{tabular}{lllllll}
\hline
System no&basic $R_{C,S}$  & mean $R_{C,S}$   & std $R_{C,S}$ &basic $d_{S}$  & mean $d_{S}$  & std $d_{S}$\\
\hline
 1&0.578 &  0.582&  0.0311&
757 &  779 & 41.8\\
2&0.623  & 0.623 & 0.0419&
 723  & 722 & 35.4\\
3& 0.575  & 0.575 & 0.0483&
 732  & 740 & 34.89\\
4& 0.782  & 0.792 & 0.0571&
 749  & 768 & 50.45\\
5& 0.609  & 0.622 & 0.099&
 578  & 590 & 46.82\\
6& 0.609  & 0.615 & 0.033&
 728  & 737 & 37.05\\
7& 0.612  & 0.628 & 0.0636&
659  & 663 & 18.43\\
8& 0.672  & 0.666 & 0.061&
 658  & 662 & 25.36\\
9& 0.703  & 0.715 & 0.049&
782  & 812 & 44.66\\
10&  0.655  & 0.651 & 0.0511&
 761  & 760 & 29.3\\
11&  0.576  & 0.551 & 0.0698&
649  & 638 & 44.94\\
12&  0.54  & 0.56 & 0.108&
 664  & 669 & 20.91\\
13&  0.694  & 0.695 & 0.043&
 692  & 695 & 22.5\\
14&  0.504  & 0.501 & 0.0312&
 726  & 731 & 35.70\\
15&  0.522  & 0.526 & 0.0594&
 620  & 621 & 24.49\\
 16& 0.8  & 0.806 & 0.0369&
 726  & 742 & 50\\
 17& 0.693  & 0.7 & 0.0341&
 717  & 724 & 24.03\\
18& 1.24  & 1.29 & 0.221&
 761  & 779 & 59.1\\
19&  0.651  & 0.63 & 0.0512&
 668  & 666 & 11.64\\
20&  0.529  & 0.5 & 0.045&
 777  & 762 & 43.25\\
 21& 0.537  & 0.534 & 0.044&
 752  & 754 & 33.42\\
\end{tabular}
\end{center}
\caption{(Supplemental) For 21 systems with $R_{C,s}>0.5$ und $d_S<800$, we re-calculate
$R_{C,s}$ and $d_S<800$
from 20 variations on concentration values within a 20\% interval. Shown are 
the original, the mean and std values.}
\label{robustness}
\end{table}
\end{document}